\documentstyle[12pt]{article}
\begin{document}

\newcommand{\be}{\begin{equation}}
\newcommand{\ee}{\end{equation}}
\newcommand{\bea}{\begin{eqnarray}}
\newcommand{\eea}{\end{eqnarray}}
\newcommand{\nn}{\nonumber}

\title{Nonperturbative Reduction of Yang-Mills Theory
    and Low Energy Effective Action}
\author{A. M. Khvedelidze \ $^a$
\thanks{Permanent address: Tbilisi Mathematical Institute,
380093, Tbilisi, Georgia}\,\,
and
\,\, H.-P. Pavel\ $^b$  \\[1cm]
$^a$ Joint Institute for Nuclear Research, Dubna, Russia\\
$^b$ Fachbereich Physik der Universit\"at Rostock,\\
              D-18051 Rostock, Germany}
\date{\empty}
\maketitle

\begin{abstract}
The method of reduction of a non-Abelian gauge theory
to the corresponding unconstrained system is exemplified
for $SU(2)$ Yang-Mills field theory.
The reduced  Hamiltonian which describes the dynamics
of the gauge invariant variables is presented in the
form of a strong coupling expansion.
The physical variables are separated into fields, which are scalars under 
spatial rotations, and rotational degrees of freedom.
It is shown how in the infrared limit an effective nonlinear sigma model 
type Lagrangian can be derived which out of the six physical fields involves 
only one of three scalar fields and two rotational fields summarized 
in a unit vector. Its possible relation to the effective Lagrangian proposed 
recently by Faddeev and Niemi is discussed.
\end{abstract}

\section{Introduction}

The perturbative reduction of non-Abelian gauge theories via gauge fixing,
which ascribes the transverse components of the gauge field as the physical
variables, is in accordance with asymptotic freedom seen in high energy
scattering processes, but is not appropriate for the description of
low energy phenomena such as confinement.
The alternative, nonperturbative, approaches to the  reduction
of gauge theories suggest different types of representations for
physical variables \cite{GoldJack}-\cite{KP} but
none of them lead directly to the gauge-invariant formulation of
the low energy problems and the question of variables
relevant to the infrared behaviour of strong theory is stil open.
The main task of the present report is to state the new
unconstrained representation for $SU(2)$ Yang-Mills system
obtained recently \cite{KP} and to
disscuss the effective field theory which follows from this
unconstrained formulation for description of infrared region.
We shall give a Hamiltonian formulation of classical $SU(2)$ Yang-Mills
field theory entirely in terms of gauge invariant quantities,
and separate six physical variables into scalars under ordinary space
rotations and into ``rotational'' degrees of freedom.
We shall obtain an effective low energy theory involving only two of the 
three rotational fields, summarized in a unit vector, and one of the tree 
scalar fields, and shall discuss its possible relation to the effective 
soliton Lagrangian proposed recently in \cite{FadNiem}.

\section{Elimination of the gauge degrees of freedom in
an adapted canonical coordinate basis}

The Hamiltonian dynamics of \(SU(2)\) gauge fields
\( A^a_{\mu}(x) \)  with the action
\be 
\label{eq:act}
{\cal S} [A] : = - \frac{1}{4}\ \int d^4x\ F^a_{\mu\nu} F^{a\mu \nu}
\quad ,\quad \quad
F^a_{\mu\nu} : = \partial_\mu A_\nu^a  -  \partial_\nu A_\mu^a
+ g \epsilon^{abc} A_\mu^b A_\nu^c~,
\ee
takes place on the definite domain of phase space spanned by the
canonical variables
\( (A^a_0,  P^a :=  \partial L/\partial (\partial_0{A}^a_0 ) \),
and  \( (A_{ai}, E_{ai}:= \partial L/\partial (\partial_0 {A}_{ai}) \).
This submanifold is defined by the three primary constraints
\be
P^a (x) = 0
\ee
and the non-Abelian Gauss law constraints
\be  \label{eq:secconstr}
\Phi_a  : = \partial_i E_{ai}+g\epsilon_{abc} A_{ci} E_{bi} = 0~,
\ee
which are first class 
\bea  \label{eq:algb1}
\{\Phi_a (x) , \Phi_b (y)\} &= & g\epsilon_{abc}\Phi_c \delta(x-y)~.
\eea

According to the Dirac prescription \cite{DiracL} the evolution of
the system is governed by the total Hamiltonian
containing three arbitrary functions \(\lambda_a (x)\)
\be \label{eq:totham}
H_T := \int d^3x\left[\frac{1}{2}\left( E_{ai}^2  + B_{ai}^2 (A) \right)
 - A^a_0 \Phi_a  + \lambda_a (x) P^a (x)
\right]~,
\ee
where
$B_{ai}(A) :=
\epsilon_{ijk}\left(\partial_j A_{ak}+{1\over 2}g
\epsilon_{abc}A_{bj} A_{ck}\right)$
is  the non-Abelian magnetic field.
The presence of these arbitrary functions reflects
the invariance of the Yang-Mills action (\ref{eq:act})
under the $SU(2)$ gauge transformations
\be  \label{eq:tran}
A_\mu \,\,\, \rightarrow \,\,\,  A_\mu^{\prime}   =
U^{-1}(x) \left( A_\mu  - {1\over g}\partial_\mu \right) U(x)~,
\ee
and leads to the problem of isolation of the gauge-invariant
functions, the observables, which are free of any constraints
and have uniquely predictable dynamics.
The reduction procedure consits of the elimination of the non-invariant,
pure gauge degrees of freedom and the formulation of the 
corresponding unconstrained system equivalent to the initial one.
Equivalence means that all observables of initial theory can be constructed
in terms of unconstrained variables and have the same unique
dynamics.

For Abelian constraints
\(\Psi_\alpha  \,\,  (\{\Psi_\alpha, \Psi_\beta\} = 0)\)
the reduction procedure can be achieved in the following two steps.
One performs a canonical transformation to new variables such that part
of the new momenta $P_\alpha$
coincide with the constraints $\Psi_\alpha $. After the projection onto the
constraint shell, i.e. putting in all expressions
$P_\alpha = 0$, the coordinates canonically conjugate to the $P_\alpha$
drop out from the physical quantities. The remaining
canonical pairs are then gauge invariant and form the basis for the
unconstrained system.
For the case of non-Abelian constraints (\ref{eq:algb1}) it is clearly
impossible to find such a canonical basis only via canonical transformation.
However one can replace the set of
non-Abelian constraints (\ref{eq:algb1}) by a new set of Abelian
constraints which describe the same constraint surface in phase space
and thus reduce the problem to the Abelian case
(see  e.g. \cite{HenTeit},\cite{GKPabel} and references therein).
This problem of Abelianization of constraints is considerably simplified
when studied in terms coordinates adapted to the action of the gauge group.
The knowledge of the $SU(2)$ gauge transformations  (\ref{eq:tran})
which leaves the Yang-Mills action (\ref{eq:act}) invariant,
directly promts us with the choice of adapted coordinates by using the
following point transformation to the new set of Lagrangian coordinates
$ \overline{Q}_j\ \ (j=1,2,3)$  and the six elements
${Q^\ast}_{ik}= {Q^\ast}_{ki}\ \ (i,k=1,2,3)$ of the positive definite
symmetric $3\times 3$ matrix $Q^\ast$
\be
\label{eq:gpottr}
A_{ai} \left(\overline{Q}, Q^{\ast} \right) :=
O_{ak}\left(\bar{Q}\right) Q^{\ast}_{\ ki}
- {1\over 2g}\epsilon_{abc} \left( O\left(\overline{Q}\right)
\partial_i O^T\left(\overline{Q}\right)\right)_{bc}\,,
\ee
where \( O(\overline{Q}) \) is an orthogonal $3\times 3$ matrix
parametrised by the $\overline{Q}_i$.
\footnote{
In the strong coupling limit the representation (\ref{eq:gpottr}) reduces
to the so-called polar representation for arbitrary quadratic matrices.
In the general case we have the additional second term and
(\ref{eq:gpottr}) has to be regarded as a set of partial
differential equations for the $\overline{Q}_i$ variables.
The uniqueness and regularity of the suggested transformation
(\ref{eq:gpottr}) depends on the boundary conditions
imposed.}
In the following we shall show that in terms of these variables the
non-Abelian Gauss law constraints (\ref{eq:secconstr})
only depend on \(\overline{Q}_i \) and its conjugated \(\overline{P}_i\)
and after Abelianization become \(\overline{P}_i = 0 \).
The unconstrained variables ${Q^\ast} $ and their conjugate \(P^\ast\) are
gauge invariant, i.e. commute with the Gauss law and all observable
quantities should depend only on ${Q^\ast} $ and \(P^\ast\).
The transformation (\ref{eq:gpottr}) induces a point canonical transformation
linear in the new canonical momenta \( P^{\ast}_{\ ik} \) and
\(\overline{P}_{i} \). Using the corresponding  generating functional
depending on the old momenta and the new coordinates,
\be
F_3 \left[ E; \ \overline{Q}, Q^{\ast}\right] :=  \int d^3z \ E_{ai}(z)
A_{ai} \left(\bar{Q}(z), Q^{\ast}(z)\right)~,
\ee
one can obtain the  transformation to new canonical momenta
\( \overline{P}_{i} \) and \( P^{\ast}_{\ ik} \)
\bea  \label{eq:mom1}
\overline{P}_j (x)& :=  & \frac{\delta F_3 }{\delta \overline{Q}_j(x)}
= - {1\over g} \Omega_{jr}\left(D_i(Q^{\ast})O^TE\right)_{ri}~,\\
\label{eq:mom2}
P^{\ast}_{\ ik}(x)& :=  &\frac{\delta F_3}{\delta Q^{\ast}_{ik}(x)}
= \frac{1}{2}\left(E^TO + O^T E \right)_{ik}~.
\eea
Here
\be
\label{Omega}
\Omega_{ji}(\overline{Q}) \, : = \,-\frac{i}{2}
\mbox{Tr}\ \left(
O^T\left(\overline{Q}\right)\frac{\partial O \left(\overline{Q}\right)}
{\partial \overline{Q}_j}
\, J_i \right),
\ee
with  the \( 3 \times 3 \) matrix generators of \(SO(3)\),
\( \left( J_i\right)_{mn} := i\epsilon_{min} \),
and the corresponding covariant derivative $D_i(Q^{\ast})$
in the adjoint  representation
\be\label{eq:newcon}
\left(D_i(Q^{\ast})\right)_{mn} := \delta_{mn}\ \partial_i
-ig \left( J^k \right)_{mn}\ Q^{\ast}_{ki}.
\ee
A straightforward calculation based on the  linear
relations  (\ref{eq:mom1}) and (\ref{eq:mom2})
between the old and the new momenta
leads to the following expression for the field strengths $E_{ai}$
in terms of the new canonical variables
\be  \label{eq:elpotn}
E_{ai} = O_{ak}\left( \overline{Q}\right) \biggl
[\,  P^{\ast}_{\ ki} +\epsilon _{kis}
 {}^\ast D^{-1}_{sl}(Q^{\ast})
\left[
\left(\Omega^{-1} \overline{P} \right)_{l} -
{\cal S}_l\,\right]\,\biggr]~.
\ee
Here  \( {}^\ast D^{-1}\) is the inverse of the matrix operator
\be
\label{DeltaQ}
{}^\ast D_{ik}(Q^{\ast}) : = -i\left(J^m D_m(Q^\ast)\right)_{ik}~,
\ee
and
\be
\label{eq:spin}
{\cal S}_k (x) := \epsilon_{klm}\left(P^\ast Q^{\ast}\right)_{lm} -
{1\over g} \partial_l P_{kl}^\ast\,.
\ee

Using the representations (\ref{eq:gpottr}) and (\ref{eq:elpotn})
one can easily convince oneself that the
variables \( {Q^\ast}\) and \({P^\ast} \) make no contribution to the
Gauss law constraints (\ref{eq:secconstr})
\be
\Phi_a : = O_{as}(\bar{Q}) \Omega^{-1}_{\ sj}
\overline{P}_j = 0~.
\label{eq:4.54}
\ee
Here and in (\ref{eq:elpotn}) we assume that the matrix
$ \Omega$ is invertible. The equivalent set of Abelian constraints is
\be
\overline{P}_a   = 0~.
\ee
They are Abelian due to the canonical structure of the new variables.

After having rewritten the model in terms of the new canonical coordinates
and after the Abelianization of the Gauss law constraints,
the construction of the unconstrained Hamiltonian system
is straightforward. In all expressions we can simply put
\(\overline{P}_i = 0\).
In particular, the Hamiltonian in terms of the unconstrained
canonical variables \(Q^\ast\) and \(P^\ast\)
can be represented by the sum of three terms
\be \label{eq:uncYME}
H[Q^\ast,P^\ast]=
\frac{1}{2} \int d^3{x}
\biggl[\,
\mbox{Tr}(P^\ast)^2 + \mbox{Tr}(B^2(Q^\ast))
\ + {1\over 2}{\vec E}^2(Q^\ast,P^\ast)
\biggr]~.
\ee
The first term is the conventional quadratic ``kinetic'' part and
the second the ``magnetic potential'' term
which is the trace of the square of the non-Abelian magnetic field
\be
 B_{sk}(Q^\ast): = \epsilon_{klm}
\left( \partial_l Q^{\ast}_{\ sm} +
\frac{g}{2}\,  \epsilon_{sbc} \, Q^{\ast}_{\ bl}Q^{\ast}_{\ cm} \right)~.
\ee
It is intersting that after the elimination of the pure gauge degrees of
freedom the magnetic field strength tensor is the commutator of the
covariant derivatives (\ref{eq:newcon})
$F_{ij} = [D_i(Q^\ast), D_j(Q^\ast)]$.

\noindent
The third nonlocal term in the Hamiltonian (\ref{eq:uncYME}) is the square
of the antisymmetric part of the electric field,
$E_{s} :=(1/2)\epsilon_{sij}{E}_{ij}$,
after projection onto the constraint surface.
It is given as the solution of the system of differential
equations\footnote{
We remark that for the solution of this equation
we need to impose boundary conditions only on the physical variables
\(Q^\ast \) in contrast to Eq. (\ref{eq:gpottr}) for which boundary
conditions only for the unphysical variables $\overline{Q}$ are needed.}
\be
\label{vecE}
{}^\ast D_{ls}(Q^\ast) E_s = g{\cal S}_l~,
\ee
with the derivative ${}^\ast D_{ls}(Q^\ast)$ defined in (\ref{DeltaQ}).
Note that the vector ${\cal S}_i(x)$, defined in (\ref{eq:spin}),
coincides up to divergence terms with the spin
density part of the Noetherian angular momentum,
$ S_i (x) := \epsilon_{ijk}A_j^aE_{ak} $,
after transformation to the new
variables and projection onto the constraint shell.
The solution ${\vec E}$ of the differential equation (\ref{vecE})
can be expanded in a $1/g$ series. The zeroth order term is
\be
\label{vecE1}
E^{(0)}_{s}=\gamma^{-1}_{sk}\epsilon_{klm}\left(P^\ast Q^{\ast}\right)_{lm}~,
\ee
with $\gamma_{ik}:= Q^{\ast}_{ik}-\delta_{ik}\mbox{Tr}(Q^{\ast})$,
and the first order term is determined as
\be
\label{vecE2}
E^{(1)}_{s} := {1\over g}
\gamma^{-1}_{sl}\left[(\mbox{rot}\ {\vec {E}}^{(0)})_l
-\partial_k P^\ast_{kl}\right]
\ee
from the zeroth order term.
The higher terms are then obtained by the simple recurrence relations
\be
\label{vecE3}
E^{(n+1)}_{s} := {1\over g}
\gamma^{-1}_{sl}(\mbox{rot}\ {\vec {E}}^{\ (n)})_s~.
\ee

\section{The unconstrained Hamiltonian in terms of scalar 
and rotational degrees of freedom}

Whereas the gauge fields transform as vectors under spatial rotations,
the unconstrained fields $Q^\ast$ and $P^\ast$
transform as second rank tensors under spatial rotations.\footnote{
Note that for a complete analysis it is necessary to investigate
the transformation properties of the field $Q^\ast$ under the whole
Poincar\'e group. We shall limit ourselves here to the isolation
of the scalars under spatial rotations and
treat $Q^\ast$ in terms of ``nonrelativistic  spin 0 and spin 2
fields'' in accordance with the conclusions obtained in the work
\cite{Faddeev79}.}
In order to separate the three fields which are invariant under spatial
rotations from the three rotational degrees of freedom we perform the
following main axis transformation of the original symmetric
$3\times 3$ matrix field $Q^\ast(x)$
\be
Q^\ast\left(\chi,\phi\right) =
R^T(\chi(x)){\cal D}\left(\phi(x)\right) R(\chi(x)),
\ee
with the orthogonal matrix $R(\chi)$ parametrized by the
three Euler angles $\chi_i=(\phi,\theta,\psi)$
and the  diagonal matrix
${\cal D}\left(\phi\right)
:=\mbox{diag}\left(\phi_1, \phi_2,\phi_3\right) $.
The main-axis-transformation of the symmetric second rank tensor field
$Q^\ast$ therefore induces a parametrization in terms of the three
rotational degrees of freedom, the Euler angles $\chi_i$, which describe
the orientation of the ``intrinsic frame'', and the diagonal elements
$\phi_i\ (i=1,2,3)$ which are scalars under spatial rotations.

The momenta $\pi_i$ and $p_{\chi_i}$, canonical conjugate
to the diagonal elements $\phi_i$ and the Euler angles
$\chi_i$, can
easily be found using the generating functional
\be
F_3 \left[ \phi_i, \chi_i; \ P^\ast \right]  : =
\int d^3 x\  \mbox{Tr}\ \left(Q^\ast P^\ast \right) =
\int d^3 x \ \mbox{Tr}\
\left( {R }^T(\chi) {\cal D}(\phi) {R }(\chi) P^\ast \right)
\ee
as
\bea
&&
\pi_i(x) = \frac{\partial {F_3}}{\partial \phi_i(x)} =
 \ \mbox{Tr} \left( P^\ast  {R}^{T} \overline{\alpha}_{i} {R}
\right),\nn\\
&&
p_{\chi_i}(x) = \frac{\partial {F_3}}{\partial \chi_{i}(x)} =
 \mbox{Tr} \left(
\frac{\partial {R}^{T} }{\partial \chi_{i}}
R \, \left[ P^\ast Q^\ast- Q^\ast P^\ast\right]
\right).
\eea
Here  $\overline{\alpha}_{i}$ are the diagonal matricis
with the elements $(\overline{\alpha}_i)_{lm}=\delta_{li}\delta_{mi}$.
Together with the  off-diagonal matricies
$(\alpha_i)_{lm}=|\epsilon_{ilm}|$
they form an orthogonal basis for symmetric  matrices.
The original physical momenta  \( P^\ast_{ik} \) can then be expressed
in terms of the new canonical variables  as
\bea \label{eq:newmom}
{P}^\ast (x) =
{R }^T(x) \left(
         \sum_{s=1}^3  \pi_s(x) \, \overline{\alpha}_{s} +
 {1\over \sqrt{2}}\sum_{s=1}^3 {\cal P}_s(x) \, {\alpha}_{s}\right)
{R }(x)~,
\eea
with
\be
{\cal P}_{i}(x):=\frac{\xi_i(x)}{\phi_j(x) - \phi_k(x)}
 \,\,\,\,\,
(cyclic\,\,\,\, permutation \,\,\, i\not=j\not= k )~,
\ee
and the $SO(3)$ left-invariant Killing vectors in terms of Euler angles
$\chi_i=(\psi,\theta,\phi)$,
\be
\xi_k(x): ={\cal M}(\theta, \psi)_{kl} p_{\chi_l}~,
\ee
with the matrix
\be \label{eq:MCmatr}
{\cal M}(\theta, \psi): =
\left (
\begin{array}{ccc}
\sin\psi/\sin\theta,     & \cos\psi,     & - \sin\psi\cot\theta     \\
-\cos\psi/\sin\theta,    & \sin\psi ,    &  \cos\psi\cot\theta      \\
            0,                  &   0 ,         &     1
\end{array}
\right ).
\ee
The antisymmetric part ${\vec E}$ of the electric field appearing in the
unconstrained Hamiltonian (\ref{eq:uncYME})
is given by the following expansion in a $1/g$ series, analogous to
(\ref{vecE1}) - (\ref{vecE3}),
\be
E_{i} = R^T_{is}\sum_{n=0}^{\infty} {\cal E}^{(n)}_{s}~,
\ee
with the zeroth order term
\be
{\cal E}^{(0)}_{i} := -\frac{\xi_i}{\phi_j+ \phi_k}
 \,\,\,\,\,
(cycl.\,\,\,\, permut. \,\,\, i\not=j\not= k )~,
\ee
the first order term given from ${\cal E}^{(0)}$ via
\be
\label{calE1}
{\cal E}^{(1)}_{i} := {1\over g}\frac{1}{\phi_j+\phi_k}
\left[\left((\nabla_{X_j}\vec{{\cal E}}^{(0)})_k
-(\nabla_{X_k}\vec{{\cal E}}^{(0)})_j
\right)-\Xi_i\right]~,
\ee
with cyclic permutations of $ i\not=j\not= k $, and
the higher order terms of the expansion determined via the recurrence
relations
\be
{\cal E}^{(n+1)}_{i} := {1\over g}\frac{1}{\phi_j+\phi_k}
\left((\nabla_{X_j}\vec{{\cal E}}^{(n)})_k
-(\nabla_{X_k}\vec{{\cal E}}^{(n)})_j\right)~.
\ee
Here the components of the covariant derivatives $\nabla_{X_k}$
in the direction of the vector field $X_i(x):= R_{ik}\partial_k$,
 \be
(\nabla_{X_i}\vec{\cal E})_b :=
X_i {\cal E}_b + \Gamma^d_{{\ }ib}{\cal E}_d~,
\ee
are determined by the connection depending only on the Euler angles
\be
\label{3dimcon}
\Gamma^b_{{\ }i a} := \left( R X_i R^T\right)_{ab}~.
\ee
It is easy to check that the connection $\Gamma^b_{{\ }i a}$ can be
written in the form
\be
\Gamma^b_{{\ }i a} = i(J^s)_{ab} ({\cal M}^{-1})_{sk} X_i \chi_k~,
\ee
using the matrix ${\cal M}$ given
in terms of the Euler angles $\chi_i=(\psi,\theta,\phi)$
in (\ref{eq:MCmatr}),
which expresses the dual nature of the Killing vectors $\xi_i$
in (\ref{eq:MCmatr}), and the Maurer-Cartan one-forms $\omega^i$ defined by
\be
RdR^T =: \omega^iJ^i, \quad \quad \quad
\omega^i = ({\cal M}^{-1})^i_kd\chi_k~.
\ee
The source terms $\Xi_k$ in (\ref{calE1}), finally, are given as
\be
\Xi_1= \Gamma^1_{{\ }2 2}(\pi_1-\pi_2)+{1\over 2}X_1\pi_1
       -\Gamma^2_{{\ }2 3}{\cal P}_2-\Gamma^1_{{\ }2 3}{\cal P}_1
       -2\Gamma^1_{{\ }1 2}{\cal P}_3+X_2{\cal P}_3
       \ +\ (2 \leftrightarrow 3)~,
\ee
and its cyclic permutaions $\Xi_2$ and $\Xi_3$.

The unconstrained Hamiltonian therefore takes the form
\be \label{eq:unch2}
H \, = \,
\frac{1}{2} \int d^3x
\left( \sum_{i=1}^3\pi_i^2
 + {1\over 2}\sum_{cycl.}{\xi^2_i\over (\phi_j-\phi_k)^2} +
{1\over 2}{\vec{\cal E}}^{\ 2} + V  \right)~,
\ee
where the potential term $V$
\be
\label{Vinhom1}
V[\phi,\chi] = \sum_{i=1}^{3} V_i[\phi,\chi]
\ee
is the sum of
\bea
\label{Vinhom2}
V_1[\phi,\chi]& = &\left(\Gamma^1_{{\ }12}(\phi_2-\phi_1)
                              -X_2 \phi_1\right)^2
+\left(\Gamma_{{\ }13}^1(\phi_3-\phi_1)
                      -X_3 \phi_1\right)^2
                   + \nn\\
&&\left(\Gamma_{{\ }23}^1\phi_3
                +\Gamma_{{\ }32}^1\phi_2-g\phi_2 \phi_3\right)^2~,
\eea
and its cyclic permutations .
We see that, via the main-axis-transformation of the symmetric second rank
tensor field $Q^\ast$, the rotational degrees of freedom, the Euler angles
$\chi$ and their canonical conjugate momenta $p_\chi$, have been isolated
from the scalars under spatial rotations, and appear in the unconstrained
Hamiltonian only via the three Killing vector fields $\xi_k$,
the connections $\Gamma$, and the derivative vectors $X_k$.

\section{The infrared limit of unconstrained $SU(2)$ gluodynamics}


From the expression (\ref{eq:unch2}) for the  unconstrained  Hamiltonian
one can analyse the classical system in the strong coupling
limit up to order $O(1/g)$. Using the leading order (\ref{calE1}) of
the $\vec{{\cal E}}$ we obtain the Hamiltonian
\be
\label{eq:strongh}
H_{S} \, = \,
\frac{1}{2}\int d^3x \left(\sum_{i=1}^3 \pi_i^2
 + \sum_{cycl.}\xi^{2}_i\frac{\phi_j^2+\phi_k^2 }{(\phi_j^2-\phi_k^2)^2}
+ V[\phi,\chi] \right)~.
\ee
For the further investigation of the low energy properties of $SU(2)$
field theory a thorough understanding of the properties of the term in
(\ref{Vinhom1}) containing no derivatives,
\be
\label{Vhomo}
V_{\rm hom}[\phi_i] = g^2[\phi_1^2\phi_2^2+\phi_2^2\phi_3^2
+\phi_3^2\phi_1^2]~,
\ee
is crucial. The classical absolute minima of energy correspond to
vanishing of the positive definite kinetic term
in the Hamiltonian (\ref{eq:strongh}). The stationary points of the
potential term (\ref{Vhomo}) are
\be
\phi_1=\phi_2=0\ ,\ \ \ \phi_3\ -\ {\rm arbitrary}~,
\ee
and its cyclic permutations.
Analysing the second order derivatives of the potential
at the stationary points one can conclude that
they form a continous line of degenerate absolute minima at zero energy.
In other words the potential has a ``valley'' of zero energy minima
along the line $\phi_1=\phi_2=0$.
They are the unconstrained analogs of the toron solutions \cite{Luescher}
representing constant Abelian field configurations with vanishing magnetic
field in the strong coupling limit.
The special point $\phi_1=\phi_2=\phi_3=0$ corresponds to the ordinary
perturbative minimum.

For the investigation of configurations of higher energy it is necessary to
include the part of the kinetic term in (\ref{eq:strongh}) containing
the angular momentum variables $\xi_i$.
Since the singular points of this term just correspond to the
absolute minima of the potential there will a competition between an
attractive and a repulsive force. At the balance point we shall have a local
minimum corresponding to a classical configuration with higher energy.

We now would like to find the effective classical field theory
to which the unconstrained theory reduces in the limit of infinite
coupling constant $g$, if we assume that the classical system spontaneously
chooses one of the classical zero energy minima of the leading order $g^2$ 
part (\ref{Vhomo}) of the potential. 
As discussed above these classical minima include apart from the perturbative 
vacuum, where all fields vanish, also field configurations with one scalar 
field attaining arbitrary values. 
Let us therefore put without loss of generality 
(explicitly breaking the cyclic symmetry)
\be
\phi_1=\phi_2=0\ ,\ \ \ \phi_3\ -\ {\rm arbitrary}~,
\ee
such that the potential (\ref{Vhomo}) vanishes.
In this case the part of the potential (\ref{Vinhom1}) containing 
derivatives takes the form
\bea
V_{\rm inh} &=&
 \phi_3(x)^2\big[(\Gamma^2_{{\ }1 3}(x))^2+(\Gamma^2_{{\ }2 3}(x))^2
           +(\Gamma^2_{{\ }3 3}(x))^2+\nonumber\\
&& \ \ \ \ \ \ \ \     +(\Gamma^3_{{\ }1 1}(x))^2+(\Gamma^3_{{\ }2 1}(x))^2
           +(\Gamma^3_{{\ }3 1}(x))^2 \big]+\nonumber\\
&& +\big[(X_1\phi_3)^2+(X_2\phi_3)^2\big]
   +2\phi_3(x)\big[\Gamma_{{\ }3 1}^3(x) X_1\phi_3
                   +\Gamma_{{\ }3 2}^3(x) X_2\phi_3\big]~.
\eea
Introducing the unit vector
\be
n_i(\phi,\theta):=R_{3i}(\phi,\theta)~,
\ee
pointing along the 3-axis of the ``intrinsic frame'', one can write
\be
V_{\rm inh}= \phi_3(x)^2 \left(\partial_i {\vec n}\right)^2
              +(\partial_i\phi_3)^2
 -(n_i \partial_i\phi_3)^2 - (n_i\partial_i n_j) \partial_j (\phi_3^2)~.
\ee
Concerning the contribution from the nonlocal term in this phase,
we obtain for the leading part of the electric fields
\be
{\cal E}^{(0)}_1= -\xi_1/\phi_3\ \ ,\ \ {\cal E}^{(0)}_2=-\xi_2/\phi_3~.
\ee
Since the third component ${\cal E}^{(0)}_3$ and ${\cal P}_3$ are singular
in the limit
$\phi_1,\phi_2\rightarrow 0$, it is necessary to have $\xi_3\rightarrow 0$.
The assumption of a definite value of $\xi_3$ is in accordance with the fact
that the potential is symmetric around the 3-axis for small $\phi_1$ and
$\phi_2$, such that the intrinsic angular momentum $\xi_3$
is conserved in the neighbourhood of this configuration.
Hence we obtain the following effective Hamiltonian up to  order $O(1/g)$
\bea
H_{\rm eff} &=& {1\over 2}\int d^3x
\bigg[\pi_3^2+{1\over \phi_3^2}(\xi_1^2+\xi_2^2) +(\partial_i\phi_3)^2
+\phi_3^2(\partial_i{\vec n})^2 \nonumber\\
&&\ \ \ \ \ \ \ \ \ \ \ \ \ \ \
 -(n_i \partial_i\phi_3)^2 - (n_i\partial_i n_j) \partial_j (\phi_3^2)
\bigg]~.
\eea
After the inverse Lagrangian transformation we obtain the corresponding
nonlinear sigma model type effective Lagrangian for  the unit vector
${\vec n}(t,{\vec x})$ coupled to the scalar field $\phi_3(t,\vec{x})$
\be
\label{Leff}
L_{\rm eff}[\phi_3,{\vec n}]= {1\over 2}\int d^3x 
\left[(\partial_\mu \phi_3^2)^2+
       \phi_3^2(\partial_\mu {\vec n})^2
      +(n_i \partial_i\phi_3)^2 
      + n_i(\partial_i n_j) \partial_j (\phi_3^2)\right]~.
\ee
In the limit of infinite coupling the unconstrained field theory in terms of
six physical fields equivalent to the original $SU(2)$ Yang-Mills theory 
in terms of the gauge fields $A_\mu^a$ reduces therefore to an effective 
classical field theory involving only one of the three scalar fields and
two of the three rotational fields summarized in the unit vector $\vec{n}$.  
Note that this nonlinear sigma model type Lagrangian
admits singular hedgehog configurations of the unit vector field $\vec{n}$.
Due to the absence of a scale at the classical level, however, these are
unstable. 
Consider for example the case of one static monopole placed at the origin,
\be
n_i:= x_i/r~,\ \ \ \phi_3=\phi_3(r)~, \ \ \ r:=\sqrt{x_1^2+x_2^2+x_3^2}~.
\ee
Minimizing its total energy $E$ 
\be
E[\phi_3]=4\pi \int dr \phi_3^2(r)
\ee 
with respect to $\phi_3(r)$ we find the classical solution $\phi_3(r)\equiv 0$.
There is no scale in the classical theory.
Only in a quantum investigation a mass scale such as a nonvanishing value for
the condensate $<0|\hat{\phi}_3^2|0>$ may appear, which might be related to 
the string tension of flux tubes directed along the unit-vector field 
${\vec n}(t,{\vec x})$. 
The singular hedgehog configurations of such string-like directed flux tubes 
might then be associated with the glueballs.
The pure quantum object $<0|\hat{\phi}_3^2|0>$ might be realized as a squeezed 
gluon condensate \cite{BlaPa}.
Note that for the case of a spatially constant condensate,
\be
<0|\hat{\phi}_3^2|0>=:2 m^2= const.~,
\ee 
the quantum effective action corresponding to (\ref{Leff}) should 
reduce to the lowest order term of the effective soliton Lagangian discussed
very recently by Faddeev and Niemi \cite{FadNiem}
\be
\label{FN}
L_{\rm eff}[{\vec n}]= m^2\int d^3x (\partial_\mu {\vec n})^2~.
\ee
As discussed in \cite{FadNiem}, for the stability of these knots 
furthermore a higher order Skyrmion-like term in the derivative expansion 
of the unit-vector field ${\vec n}(t,{\vec x})$ is necessary.
To obtain it from the corresponding higher order terms in the strong coupling 
expansion of the unconstrained Hamiltonian (\ref{eq:unch2}) is under present 
investigation.

\section{Concluding remarks}

Following the Dirac formalism for constrained Hamiltonian systems
we have formulated classical $SU(2)$ Yang-Mills gauge theory
entirely in terms of unconstrained gauge invariant local fields.
All transformations which have been used,
canonical transformations and the Abelianization of the constraints,
maintain the canonical structures of the generalized Hamiltonian
dynamics.
We identify the unconstrained fields with symmetric positive
definite second rank tensor fields $Q^\ast$ and $P^\ast$
 under spatial rotations. The three scalar fields are separated
from the three rotational degrees of freedom
via the main-axis transformation of the field $Q^\ast$.
Our unconstrained representation of the Hamiltonian furthermore
allows us to derive an effective low energy Lagrangian for the rotational 
degrees of freedom coupled to one of the scalar fields
suggested by the form of the classical potential
in the strong coupling limit. The dynamics
of the rotational variables in this limit is summarized
by the unit vector describing the orientation of the intrinsic
frame. Due to the absence of a scale in the classical theory the
singular hedgehog configurations of the unit vector field is found to be 
unstable classically. Only in a quantum treatment,
which is under present investigation, a nonvanishing value for the 
vacuum expectation value for one of the three scalar field operators, 
and hence a mass scale, can occur.  
For the case of a spatially constant scalar quantum condensate
we expect to obtain the first term of a derivative expansion proposed recently 
by Faddeev and Niemi \cite{FadNiem}. As shown in their work
such a soliton Lagragian allows for stable massive knotlike configurations
which might be related to glueballs. For the stability of the knots
higher order terms in the derivative expansion, such as the Skyrme type
fourth order term in \cite{FadNiem}, are necessary. Their derivation
in the framework of the unconstrained theory, proposed in this paper, 
is also under investigation.


The work of A.M.K. was partially supported  by the Russian Foundation for
Basic Research under grant No. 98-01-00101.
H.-P. P. acknowledges support by the Deutsche Forschungsgemeinschaft
under grant No. Ro 905/11-2 and by the Heisenberg-Landau program for
providing a grant.


\end{document}